\documentclass{article}
\usepackage{url}
\usepackage{graphicx}
\usepackage{epsfig}
\usepackage{amsmath}

\begin{document}
%
\title{Agent mental models and Bayesian rules as a tool to create opinion dynamics models}
%
%
\author{Andr\'e C. R. Martins \\  
%
%
%
	GRIFE -- EACH -- Universidade de S\~ao Paulo,\\
	Rua Arlindo B\'etio, 1000, 03828--000,  S\~ao Paulo, Brazil
}

\maketitle              

\begin{abstract}
Traditional models of opinion dynamics provide a simple approach to understanding human behavior in basic social scenarios. However, when it comes to issues such as polarization and extremism, we require a more nuanced understanding of human biases and cognitive tendencies. In this paper, we propose an approach to modeling opinion dynamics by integrating mental models and assumptions of individuals agents using Bayesian-inspired methods. By exploring the relationship between human rationality and Bayesian theory, we demonstrate the efficacy of these methods in describing how opinions evolve. Our analysis leverages the Continuous Opinions and Discrete Actions (CODA) model, applying Bayesian-inspired rules to account for key human behaviors such as confirmation bias, motivated reasoning, and our reluctance to change opinions. Through this, we obtain update rules that offer deeper insights into the dynamics of extreme opinions. Our work sheds light on the role of human biases in shaping opinion dynamics and highlights the potential of Bayesian-inspired modeling to provide more accurate predictions of real-world scenarios.
\end{abstract}
\section{Introduction: The need for general methods}

Opinion dynamics models~\cite{castellanoetal07,galam12a,latane81a,galametal82,galammoscovici91,sznajd00,deffuantetal00,martins08a,martins12b} is a fascinating area of research that seeks to understand how opinions spread through society. A plethora of models have been developed to describe this process, ranging from simple to complex, and covering topics such as the formation of consensus\cite{schawe2021bridges}, the emergence of polarization \cite{DiMaggio1996,Baldassarri2008,Taber2009,Dreyer2019}, the different ways we can define it \cite{Bramson2017}, and spread of extreme opinions \cite{deffuantetal02a,amblarddeffuant04,galam05,weisbuchetal05,franksetal08a,martins08b,Li2013,Parsegov2017,Amelkin2017}.
Extremism can be defined as the end of a range over continuous variable ~\cite{deffuantetal00,hegselmannkrause02}, or as inflexibles who do not change their minds\cite{galamjacobs07}, or using mixed models \cite{martins08a,martins12b,martinsgalam13a}. To explore the problem of extremism in the real world, not only opinions matter \cite{Martins2022a} and we must also consider actions as part of what defines an extremist \cite{tileaga06a,bafumiherron10a}.

However, despite the wealth of knowledge already gathered, there are still many aspects of opinion dynamics that require greater attention. Community efforts are necessary to fill gaps in research and promote progress in the field \cite{Sobkowicz2020}. 
Currently, most models are only comparable to similar implementations, with a lack of translation between different types. While attempts to propose general frameworks and universal formulas do exist \cite{martins12b,Boettcher2017,Galam2020}, they are, so far, isolated efforts. To achieve greater understanding, we need to explore how different models relate to each other \cite{martins13c} and develop methods to incorporate new effects and assumptions.

To gain a deeper understanding of the spread of polarization and extremism, we must also consider actions, not just opinions. Incorporating decision-making and behavioral aspects is crucial in modeling opinions \cite{KowalskaPyzalska2014,MuellerHansen2017}, as it allows for a more accurate depiction of how individuals perceive and react to complex information \cite{Haghtalab2021}. One promising avenue of exploration is the use of Bayesian-inspired models \cite{martins12b}.

Bayesian rules to model opinions have been introduced both in the opinion dynamics community, as extensions of the Continuous Opinions and Discrete Actions (CODA) \cite{martins08a,martins12b}  and similar opinion models \cite{martins12b,martins08b,martinsgalam13a,martins13c,martins08c,martinspereira08a,martinskuba09a,vicenteetal08b,sietal10a,Si2010,martins10b,martins10a,Deng2013,martins13b,Diao2014,luoaetal14a,Caticha2015,martins15b,Lu2015,martins16a,Chowdhury2016,Cheng2016,Huang2016,Garcia2017,Sun2017,Sobkowicz2018,Lee2018,Garcia2018,Tang2019,Martins2019,Martins2020,LeonMedina2020,Maciel2020,Fang2020,Martins2022a}, by the use of Bayesian belief networks \cite{Sun2013}, as well as, independently, in models associated to economical reasoning \cite{orleans95,Rabin1999,Andreoni2012,Nishi2013,eguiluzetal15a,Wang2016}. Despite their popularity, there are two aspects of Bayesian-inspired models for opinion dynamics that have not been properly debated so far. They are how to turn assumptions on how the agents reason on dynamical model equations and the problem of the relationship between Bayesianism and rationality.

In this paper, I will explore both aspects of the problem. First, I will provide a brief explanation on why the use of Bayesian-inspired rules is both supported by experimental evidence \cite{Knill2004,martins06,tenenbaumetal07,Tenenbaum2011,eguiluzetal15a} and not the same as assuming rationality \cite{Martins2020inpress,Martins2020a}. And, to illustrate how Bayesian rules can be used in a general problem, I will explain how we can create mental models for the agents. And we will see how models can include any kind of bias and bounded rationality effects \cite{simon56,selten01}, and turn those assumptions into update rules. More exactly, I will show how to introduce agents who distrust opinions opposed either to a certain choice (a direct bias), or against their current preference. Update rules for both kinds of agents will be calculated and the effects of those biases on how extreme the position of the agents become will be studied.

\section{Bayesian models and rationality}

Bayesian methods are one of our golden standards for rationality and inductive arguments \cite{Martins2020a}. If we start with simple rules about how induction about plausibilities must be performed, we can show that plausibilities should be updated using Bayes theorem \cite{cox61a,jaynes03}. The same theorem can be obtained from other axioms, such as maximization of entropy \cite{catichagiffin07a} or the much weaker basis of ``dutch books''. However, using those ideas to describe how people reason has been considered problematic \cite{Eberhardt2011}. On the other hand, Bayesian ideas can be used both in a hard way, strictly following its rules, or as a soft version, where its basic ideas are used to represent aspects such as updating subjective opinions \cite{Elqayam2013}. That poses the question of whether Bayesian rules can be used to describe our reasoning well. Of course, we should also ask about the requirements we impose from our rationality models. Even the definition of bounded rationality can be challenged, as it assumes there is someone to judge if any behavior is entirely rational \cite{Chater2018}. Indeed, using Bayesian methods with perfection is impossible as they require infinite abilities \cite{Martins2020a}. We can only approximate them by considering a limited set of possibilities, and that is compatible with how our brains work.

There is good evidence that we reason in ways that are similar to Bayesian methods \cite{Knill2004,martins06,tenenbaumetal07,Tenenbaum2011,eguiluzetal15a}. But, if we want to use it in mathematical and computational models, we need to go further than just similarity. Indeed, it is clear that humans are not perfectly rational nor good at statistics. We sometimes fail at easy problems \cite{watsonjohnsonlaird72a,tversykahneman83a} and we tend to be too confident about our mental skills \cite{oskamp65a}. At first, experiments about our cognitive abilities seemed to point at a remarkable amount of incompetence.

But that is not the whole story. When we look closer, some of our mistakes are not as serious as they look. While we are not good abstract logicians, when the same problems are presented associated with normal day-to-day circumstances, we answer them correctly \cite{johnsonlairdetal72a}. And there is evidence that many of our mistakes can be described as the use of reasonable heuristics  \cite{gigerenzertodd00a}, short-cuts that allow us to arrive at answers fast and with less effort \cite{tverskykahneman73a,gigerenzergoldstein96}. As simplified rules, heuristics fail under some circumstances. If we go looking for those cases, we will undoubtedly find them. But they are not a sign of complete incompetence.

That does not explain our overconfidence problems, of course. But we have also observed that our reasoning skills might not have evolved to find the best answers, even if we can use them for that purpose. Instead, humans show a tendency to defend their identity-defining beliefs \cite{kahan13a,Kahan2017b}. More than that, our ancestors had solid reasons to be good at fitting inside their groups and, if possible, ascend socially inside those. Our reasoning and argumentative skills were more valuable from an evolutionary perspective if they worked towards agreeing or convincing our peers. Group belonging mattered more than being right about things that would not affect direct survival \cite{merciersperber11a,Mercier2017}. Being sure and defending ideas from our social group would have been more important than looking for better but unpopular answers.

And there is one more issue. In many laboratory problems where we observed humans make mistakes, scientists used questions that would never appear in real life \cite{Martins2020inpress}. Take the case of the observation of weighting functions, that we seem to alter probability values that we hear \cite{kahnemantversky79}. Using changed values might seem to serve no purpose at first. However, the scientists who performed those experiments assumed the values they presented were known with certainty. But there is no such certainty in real life. If someone tells you there is a 30\% chance of rain tomorrow, even if based on very well-calibrated models, you know that is an estimate. As with any estimates, at best, the actual value is close to 30\%. A Bayesian assessment of the problem would combine our previous estimate of rain with the information about the forecast to obtain a final opinion. If we do that with the many experiments that showed we use probability values wrong, we see that our behavior is compatible with Bayesian rules. The observed changes match reasonable assumptions for everyday situations when we hear uncertain estimates \cite{martins06}. Doing that would be wrong in artificial cases, such as those in the laboratory experiments, where there is no (or very little) uncertainty about the probability values. Even our tendency to change our opinions far less than we should, called conservatism (no relationship to politics implied in the technical term) \cite{edwards68} can easily be explained. We just need to include in a Bayesian model a tendency to skepticism about the data we hear \cite{martins06}. Our brains might just have heuristics that mimick Bayesian estimates for an uncertain and unreliable world.

That is, we are not perfect, but our bounded abilities are not those of incompetents. We make reasonable approximations. We are motivated reasoners, more interested in defending ideas than looking for better answers. Given the right preferences, that can even be described as rational, despite ethical considerations. Even when it seemed we were making mistakes, we might have been behaving closer to Bayesian rules than it was initially assumed.

\section{Update rules from the agent mental models}
 
 We are not perfectly rational, but we can still be described by Bayesian rules. Therefore, it makes sense to try Bayesian methods as a way to represent human opinions. Thus, the next question we must answer is how we can include our biases and cognitive characteristics in our models. To do that, we must consider how the agents think, what they show to others, and what they expect to see from their neighbors. That is, we need to describe their mental models. But, first, it makes sense to ask how Bayesian methods work.
 
 \subsection{A very short introduction to Bayesian methods}
 
 While Bayesian statistics, done correctly, can become complicated fast, it is based on an elementary, almost trivial basis, the Bayes Theorem. It works like this. We have an issue we want to learn about, and we represent it by a variable random $X$, where each $x$ represents one possible value. Here, $x$ can be a quantity, but it can also be nothing more than a label. We start with a probability distribution, our initial guess, on how likely each possible $x$ is, represented by a probability distribution $f(x)$, called the prior opinion. Once we observe data $D$, we must change our opinions on $x$. To do that, we need to know, for each possible value $x$, how likely it would have been that we would observe $D$. That is, we need the likelihood, $f(D|x)$. From that, calculating the posterior estimate $f(x|D)$ is done just by a simple multiplication $f(x|D) \propto f(x)f(D|x)$. The proportionality constant is calculated by imposing that the final distribution must add (or integrate) to one. Everything in Bayesian methods is a consequence of that update rule and considerations on how to use it. The basic idea, already using an opinion dynamics problem, is represented in Figure
 
 \begin{figure}[h]
 	\centering
 	\includegraphics[width=0.75\textwidth]{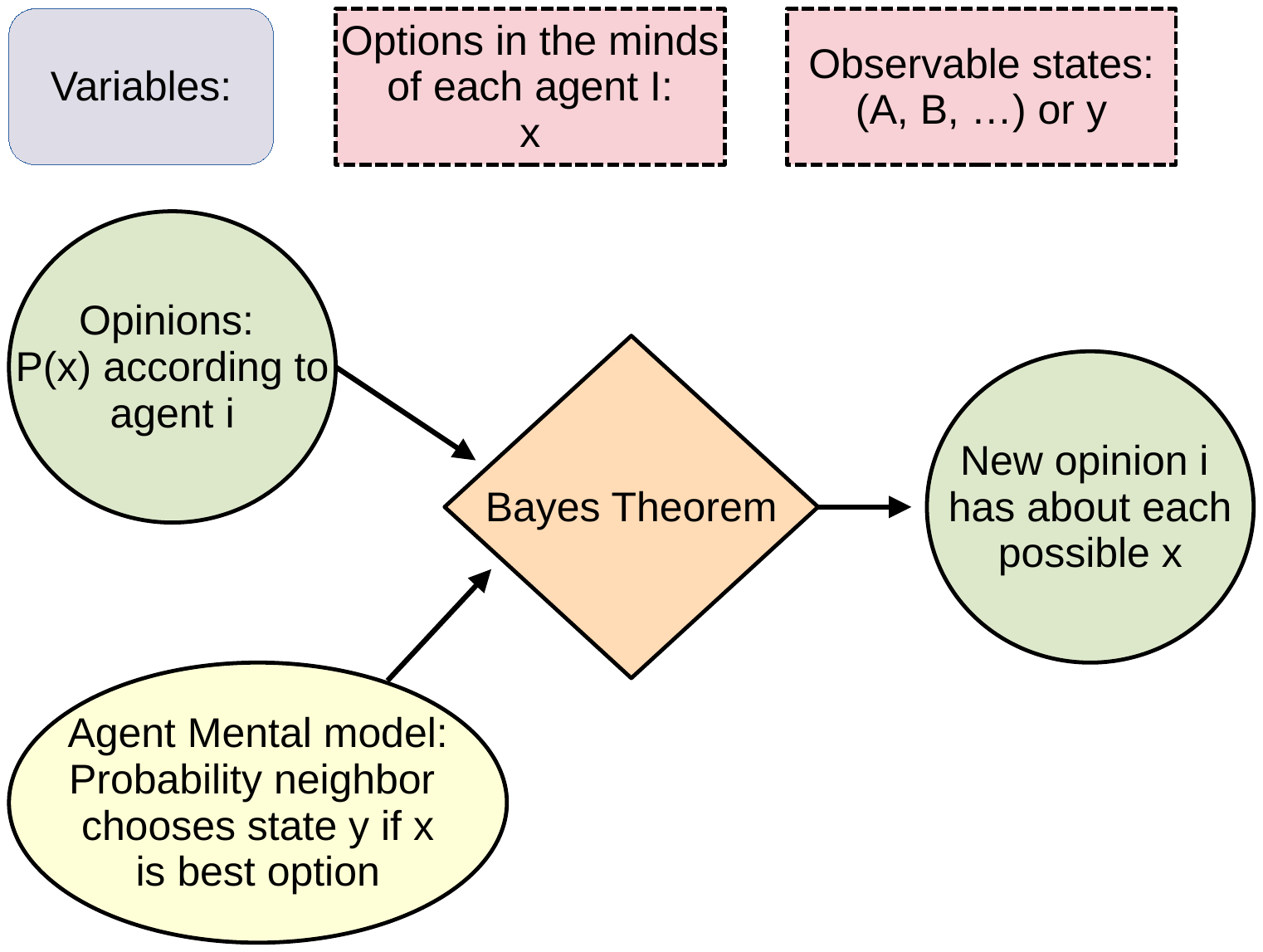}
 	\caption{Schematics for the general use of Bayes theorem as a tool for creating opinion update equations, highlighting the role of agent mental models for what others communicate to them (likelihoods).    }
 	\label{fig:scheme}
 \end{figure}

 To illustrate how it is done in practice, let us look at the demonstration of the CODA model rules. In CODA, the agents try to decide between two possible choices, $A$ or $B$ (sometimes represented as values of a spin, $+1$ or $-1$). Each agent $i$ has, a time $t$, a probability opinion $p_i(t)$ that $A$ is better than $B$ (and, $1-p_i(t)$ that $B$ is better).
 But, instead of expressing their probabilistic opinion, they only show their neighbors the option they considered more likely to be better.
 They also assume their neighbors have a larger than 50\% chance $\alpha$ to pick the best option. In principle, there could be assymetric different chances, $\alpha$ to choose $A$ when $A$ is better and $\beta$ to choose $B$ when $B$ is better. As a first example, we will assume the $\alpha=\beta$ symmetry here. From Bayes rule, we get the update model for $p_i(t+1) \propto p_i(t)\alpha$ and, similarly, $1-p_i(t+1) \propto (1- p_i(t))(1-\alpha)$. As the probabilities must add one, we divide by their sum and get the update rule
 \begin{equation}
 	p_i(t+1) = \frac{p_i(t)\alpha}{p_i(t)\alpha+(1- p_i(t))(1-\alpha)}.
 \end{equation}
 At this step, we have an update rule and we could use it as it is. In this case, however, it is trivial to make a change of variables that will provide us a much more computationally-light rule. The update rule becomes much simpler if we make the transformation
 \begin{equation}\label{eq:nu}
 \nu=\ln(\frac{p}{1-p}). 
 \end{equation}
 The denominators cancel and we get
 \begin{equation}\label{eq:CODA}
 	\nu(t+1) = \nu(t) \pm C,
 \end{equation}
 where $C=\ln(\frac{\alpha}{1-\alpha})$ and the sign on the sum depends on whether the neighbor prefers $A$ ($+$) or $B$ ($-$). We can get an even simpler model by renormalizing $\nu$ and making $C=1$. 
 
 And that is it. We start from the initial opinion, use the Bayes theorem and we get an update rule. In this case, the final rule is to add one when an neighbor prefer $A$ and subtract one when it prefers $B$, flipping opinions at $\nu=0$.

 \subsection{Agent communication rules and their mental assumptions}
 
 There were two major assumptions in the CODA model. One was how agents communicate. While they have a probabilistic estimate of which option is better, everyone else observes only their best estimate. The second assumption is the mental model of the agents. They think their neighbors are more likely than not to pick the best choice. And all those neighbors are assumed to have the same chance, $\alpha>0.5$, to get it right. Of course, we could introduce some agents that think their neighbors are more likely to be wrong, that is, $\alpha<0.5$. If we do that, we have just included contrarians \cite{martinskuba09a} in our model, that is, agents who tend to disagreement \cite{galam04}.

Making the model assumption explicit makes it easier to investigate what happens if agents behave or act differently. For example, we could have a situation where agents look for the best choice between $A$ and $B$. Still, they communicate their probability estimates that $A$ is the better choice, $p_i(t)$. In that case, while we can keep the probability $p_i(t)$ as a measure of the opinion of the agents, those agents no longer state a binary choice but a continuous specific probability value. 

\subsubsection{Changing what is communicated}

Mental models become crucial, including which question the agents are trying to answer. Agents may still want to determine which is better,$ A $ or $ B $, sharing information on their uncertainty. Or they might see the probability values as an ideal mixture. They might accept, for example, that the perfect position is 60\% of $A$ and 40\% of $B$. First, let us consider the case where they just want to pick the best choice. 

In that case, $p_i(t)$ is still a simple value associated with $A$, while, trivially, $1-p_i(t)$ gives us the probability $B$ is better. But we must evaluate the chance other agents will have any continuous value for $p_j(t)$ if $A$ is better (or if $B$ is). That is, we need a distribution probability of probabilities to represent the agent's mental model. In mathematical terms, we need a model that says, assuming $A$ is better, how likely the neighbor $j$ would have an opinion $p_j$ if $A$ is better, $f(p_j|A)$. Of course, we also need $f(p_j|B)$, but in many situations of interest, that can be obtained by symmetry assumptions. This model was implemented originally by assuming $f(p_j|A)$ was a Beta function  $Be(p_j|\alpha,\beta, A)$, that is
\[
f(p_j|A)=Be(p_j|\alpha,\beta, A)=\frac{1}{B(\alpha,\beta)}p_{j}^{\alpha-1}(1-p_{j})^{\beta-1}
\],
where $B(\alpha,\beta)$ is obtained from Gamma functions by
\[
B(\alpha,\beta)=\frac{\Gamma(\alpha)\Gamma(\beta)}{\Gamma(\alpha+\beta)}.
\] 
Here, $\alpha$ and $\beta$ are the traditional parameters the Beta function. Interestingly, the update rule can once more be defined in terms of the log-odds variable $\nu=\ln(\frac{p}{1-p})$, and that leads once more to an additive model. However, the term to be added depends on the probability $p_j$ communicated by $j$, and as the agents become more certain, the size of the additive term explodes. Consequently, extreme opinions become much stronger than the already extreme opinions in the original CODA model \cite{martins16a,Martins2020c}.
 
 \subsubsection{Changing the mental variables}
 
 While in that example, we did need a new likelihood, the Beta function that tells us how likely agents think others will provide each answer, that need was caused by a change in the communication. But it is not only what is communicated that can be changed. Inner assumptions agents make can also be changed, including the questions they want to answer. That is, we can have very different assumptions in their mental model. 
 
 Assume that, instead of having ``wisher'' agents looking for the best option between $A$ and $B$, each ``mixer'' agent has an estimate about the best mixture between $A$ and $B$. In this case, $p_i$ is the percentage of $A$ in the best blend of two options, and, as such, each value $0\leq p_i \leq 1$ must have a probability. We need probability densities $f(p)$ as prior and posterior opinion distributions. The easier way to implement such a model is to look for conjugate distributions \cite{ohagan94a}. Conjugate distributions correspond to those cases where, for a certain likelihood, the prior and the posterior will be represented by the same function. In that case, update rules can just update the parameters of that distribution function and not complete distributions. However, this is not the general case for a random application of the Bayes rule. As we build more detailed models, finding conjugate distributions might not be possible for a given mental model. 
 
 Luckily, for the more straightforward cases we are interested in, conjugate options do exist. Before proceeding to a model, we need to decide how the agents communicate. Here, once more, we can have a discrete communication, where each agent just tells others which option, $A$ or $B$ should appear in a more significant amount, or communication could include the average estimate for the proportion $p$ of $A$, $E[p]$. 
 
 In the first case, with discrete communication, we can find a natural conjugate family by using Beta distributions for the opinions and a Binomial distribution for the likelihood, that is, the chance that a neighbor will choose $A$ or $B$ depending on its average estimate. Of course, other choices of distributions would be possible, and they correspond to a similar thought structure but with different dynamics. Under the Binomial-Beta option, interestingly, the dynamics of the preferred choice, given the more straightforward choice of proper functions for the prior and the likelihood, mimics the original CODA dynamics. However, while the evolution of the preferred option in the mixture is the same, the probability (or proportion) values never go to the same extreme values \cite{martins16a}.
 
 \subsubsection{Other mental models already explored}
 
 Other variations are possible and have been explored. An initial approach to trust was implemented in a fully connected setting by adding one assumption to the agent mental model \cite{martins13b}. Instead of assuming that every other agent had a probability $\alpha$ to get the best answer, each agent assumed there were two types of agents. Agent $i$ assumed there was a probability $\tau_{ij}$ that agent $j$  was a reliable source who would pick the best option with chance $\alpha$. But other agents could also be untrustworthy (or useless) and pick the best option with probability $\mu$ so that $\mu<\alpha$, possibly even 0.5 or lower. That is, if $A$ was the best choice, instead of a chance $\alpha$ neighbor $j$ would prefer $A$, there was a chance given by $\alpha\tau_{ij}+ \mu(1-\tau_{ij})$. Applying Bayes theorem with this new likelihood led to update rules for $p_i$ and $\tau_{ij}$. Each agent updated both its opinions on whether $A$ or $B$ would be better. And it also changed its estimates about the trustworthiness of the observed agent. The update rule could not be simplified by a transformation of variables because no exact way to uncouple the evolution of the opinion and how much agents trusted their neighbors was found.
 
 A similar idea was used for a Bayesian version for continuous communication and the ``mixer''-type of agents. That model \cite{martins08c}led to an evolution of opinions qualitatively equivalent to what we observe in Bounded Confidence models \cite{deffuantetal00,hegselmannkrause02}. That continuous model was later extended to study the problem of several independent issues when agents adjusted their trust based not only on the debated subject but also on their neighbor's positions on other matters \cite{Maciel2020}. Interestingly, that did cause opinions to become more clustered and aligned, similar to the irrational consistency we observe in humans  \cite{jervis76a}.
 
 Even the agent's influence on its neighbors can be used for their mental models. That was introduced as a simple version by assuming that there were different chances $a$ and $c$ that a neighbor would prefer $A$ in the case $a$ was indeed better, depending on whether the observing agent selected $A$ or not \cite{martins13c}. That actually weakened the reinforcement effects of agreement, as the other agent could think $A$ was better not because it was but because the observer also thought that. In the limit of strong influence, the dynamics of the voter model \cite{cliffordsudbury73,holleyliggett75} -- or other types of discrete models, such as majority \cite{galam2003a,galam06b} or Sznajd \cite{sznajd00} rules, depending on the interaction rules -- was recovered. That shows that Bayesian-inspired models are much more general than the traditional discrete versions.Other variations are possible and have been explored. An initial approach to trust was implemented in a fully connected setting by adding one assumption to the agent mental model \cite{martins13b}. Instead of assuming that every other agent had a probability $\alpha$ to get the best answer, each agent assumed there were two types of agents. Agent $i$ assumed there was a probability $\tau_{ij}$ that agent $j$  was a reliable source who would pick the best option with chance $\alpha$. But other agents could also be untrustworthy (or useless) and pick the best option with probability $\mu$ so that $\mu<\alpha$, possibly even 0.5 or lower. That is, if $A$ was the best choice, instead of a chance $\alpha$ neighbor $j$ would prefer $A$, there was a chance given by $\alpha\tau_{ij}+ \mu(1-\tau_{ij})$. Applying Bayes theorem with this new likelihood led to update rules for $p_i$ and $\tau_{ij}$. Each agent updated both its opinions on whether $A$ or $B$ would be better. And it also changed its estimates about the trustworthiness of the observed agent. The update rule could not be simplified by a transformation of variables because no exact way to uncouple the evolution of the opinion and how much agents trusted their neighbors was found.
 
 A similar idea was used for a Bayesian version for continuous communication and the ``mixer''-type of agents. That model \cite{martins08c}led to an evolution of opinions qualitatively equivalent to what we observe in Bounded Confidence models \cite{deffuantetal00,hegselmannkrause02}. That continuous model was later extended to study the problem of several independent issues when agents adjusted their trust based not only on the debated subject but also on their neighbor's positions on other matters \cite{Maciel2020}. Interestingly, that did cause opinions to become more clustered and aligned, similar to the irrational consistency we observe in humans  \cite{jervis76a}.
 
 Even the agent's influence on its neighbors can be used for their mental models. That was introduced as a simple version by assuming that there were different chances $a$ and $c$ that a neighbor would prefer $A$ in the case $a$ was indeed better, depending on whether the observing agent selected $A$ or not \cite{martins13c}. That actually weakened the reinforcement effects of agreement, as the other agent could think $A$ was better not because it was but because the observer also thought that. In the limit of strong influence, the dynamics of the voter model \cite{cliffordsudbury73,holleyliggett75} -- or other types of discrete models, such as majority \cite{galam2003a,galam06b} or Sznajd \cite{sznajd00} rules, depending on the interaction rules -- was recovered. That shows that Bayesian-inspired models are much more general than the traditional discrete versions.
 
 \subsection{Introducing other behavioral questions}
 
 Bayesian rules can help us explain how humans reason \cite{martins06,Martins2020inpress}. And we just saw a few examples of introducing extra details in the agent mental models so that new assumptions can be included. In this subsection, I will discuss how we can move ahead and model some biases we observe in human behavior by applying those concepts to create an original model for a specific human tendency.

Let us start considering an easy one, confirmation bias \cite{nickerson98a}, as that does not even need new mental models. Confirmation bias is simply our tendency to look for information from sources who agree with us. As such, it can be better modeled by introducing rules that reconnect the influence network so that agents will more likely be surrounded by those who agree with them. The co-evolution of the CODA model over a network that evolved based on the agreement or disagreement of the agents and their physical location plus thermal noise was studied. Depending on the noise and the strength of the agreement term in the network rewiring, the tendency to polarization and confirmation bias was apparent \cite{Martins2019b}.

Motivated reasoning \cite{kahan13a}, on the other hand, is not only about who we learn from but how we interpret information depending on whether we agree with it or not. That can be implemented in more than one way. A simple version is an approach where trust was introduced in the CODA model \cite{martins13b}. In that model, depending on the initial conditions, as agents become more confident about their estimates, they would eventually distrust those who disagreed with them, even when they met for the first time.

	\subsubsection{Direct bias} 

 	Of course, there are other possibilities, including more heavy-handed approaches.  For example, agents might think being untrustworthy is associated with one of the two options. Instead of a trust matrix signaling how much each agent $i$ trusts each agent $j$, we can introduce trust based on the possible choices. For two options, $A$ and $B$, we can assume that each agent has a prior preference. Each agent believes that untrustworthy people will defend only the side the agent is biased against. That can be represented by a small addition to the CODA model. Assume agent $i$ prefers $A$, and it thinks that people would only go wrong to defend $B$. One way to describe that is to assume that there is a proportion $\lambda$ of reasonable agents who behave like CODA. They pick the better alternative with chance $\alpha>0.5$. However, the remaining $1-\lambda$ choose $B$ more often, regardless of whether it is true or not, with probability $\beta>0.5$. That is, for agents biased towards $A$, the chance a neighbor would choose $A$, represented by $CA$, if $A$ (or $B$) is better is given by the equations
\[
P(CA|A)= \lambda\alpha + (1-\lambda)(1-\beta),
\] 
and
\[
P(CA|B)= \lambda(1-\alpha) + (1-\lambda)(1-\beta).
\]
Also,
\[
P(CB|A)= \lambda(1-\alpha) + (1-\lambda)\beta
\] 
and
\[
P(CB|B)= \lambda\alpha + (1-\lambda)\beta.
\]

In principle, we could introduce an update rule for both the probability $p_i$ that $A$ is better and $\lambda$. However, for this exercise, let us assume there is an initial fixed value for $\lambda$. For example, to illustrate how this bias can change the CODA model, the agents can suppose that a majority of honest people are given by $\lambda=0.8$. Let us also assume that honest people get the better answer with a chance of $\alpha=0.6$, while biased people provide their wrong estimate of $B$ with a probability of $\beta=0.9$ That makes  $P(CA|A)=0.5$, $P(CA|B)=0.34$, $P(CB|A)=0.5$ and $P(CB|B)=0.66$. So, if agent $i$ observes someone who prefers $A$, it will update its opinion by $
p_i(t+1)=\frac{p_i(t)0.5}{p_i(t)0.5+(1-p_i)0.34}
$. That is, for the CODA transformed variable $\nu = \ln(\frac{p}{1-p})$, we will have $\nu_i(t+1)=\nu_i(t)+\ln(\frac{0.5}{0.34})\approx \nu_i(t)+ 0.386$. On the other hand, if $B$ is observed, the update rule will provide (again for an agent biased towards $A$), $\nu_i(t+1)=\nu_i(t)+\ln(\frac{0.5}{0.66})\approx \nu_i(t)- 0.278$. That means that steps in favor of $A$ will be larger than those of $B$ for such an agent. While that agent can be convinced by a majority, the agent will move towards its preference if there is a tie in the neighbors. Depending on the exact values of the parameters, the ratio between step sides can become larger.

We will assume a renormalization of the additive term to implement this bias, as it is normally done in CODA applications. Assuming agent $i$ is biased in favor of $A$ (the equations for the case where $i$ is biased in favor of $B$ are symmetric and will not be printed here), we have, for the variable $\nu_i(t)$, as defined in Equation~\ref{eq:nu}, when the neighbor also prefers $A$
\begin{equation}\label{eq:agree}
	\nu_i(t+1)= \nu_i(t) + \ln \left[ \frac{\lambda\alpha + (1 - \lambda)(1 - \beta)}{\lambda(1 - \alpha) + (1 - \lambda)(1 - \beta)}   \right].
\end{equation}
When the neighbor prefers $B$, we have
\begin{equation}\label{eq:disagree}
	\nu_i(t+1)= \nu_i(t) + \ln \left[ \frac{\lambda(1-\alpha) + (1 - \lambda)\beta }{ \lambda\alpha + (1 - \lambda)\beta}   \right].
\end{equation}
These equations trivially revert to the standard CODA model in Equation~\ref{eq:CODA} when all agents are considered honest, that is, when $\lambda=1$. For ease of further manipulations, we can define the size of the steps in both Equations~\ref{eq:agree} and~\ref{eq:disagree} as 
\[
SA = \ln \left[ \frac{\lambda\alpha + (1 - \lambda)(1 - \beta)}{\lambda(1 - \alpha) + (1 - \lambda)(1 - \beta)}   \right]
\]
\[
SD = \ln \left[ \frac{\lambda(1-\alpha) + (1 - \lambda)\beta }{ \lambda\alpha + (1 - \lambda)\beta}   \right].
\]

The first question we can ask about those steps is how they relate to each other and the original step size $C=\ln(\frac{\alpha}{1-\alpha})$. Simple algebraic manipulations show that  $SA < C$ and that $SD > -C$, as long as $\alpha>0$ in both cases. As $SD<0$, both size steps are smaller. That was to be expected. Introducing a chance the neighbor might not know what it is talking about should, indeed, decrease the information content of its opinion.

\begin{figure}
	\centering
	\begin{tabular}{c}
		\includegraphics[width=0.85\textwidth]{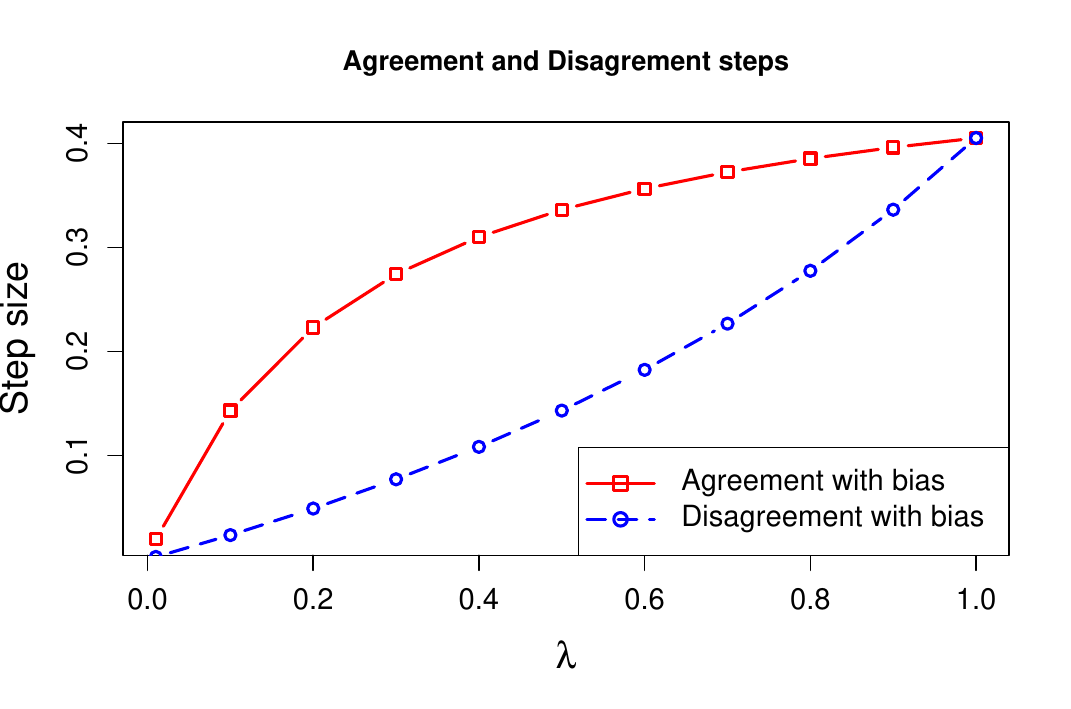}\\
		\includegraphics[width=0.85\textwidth]{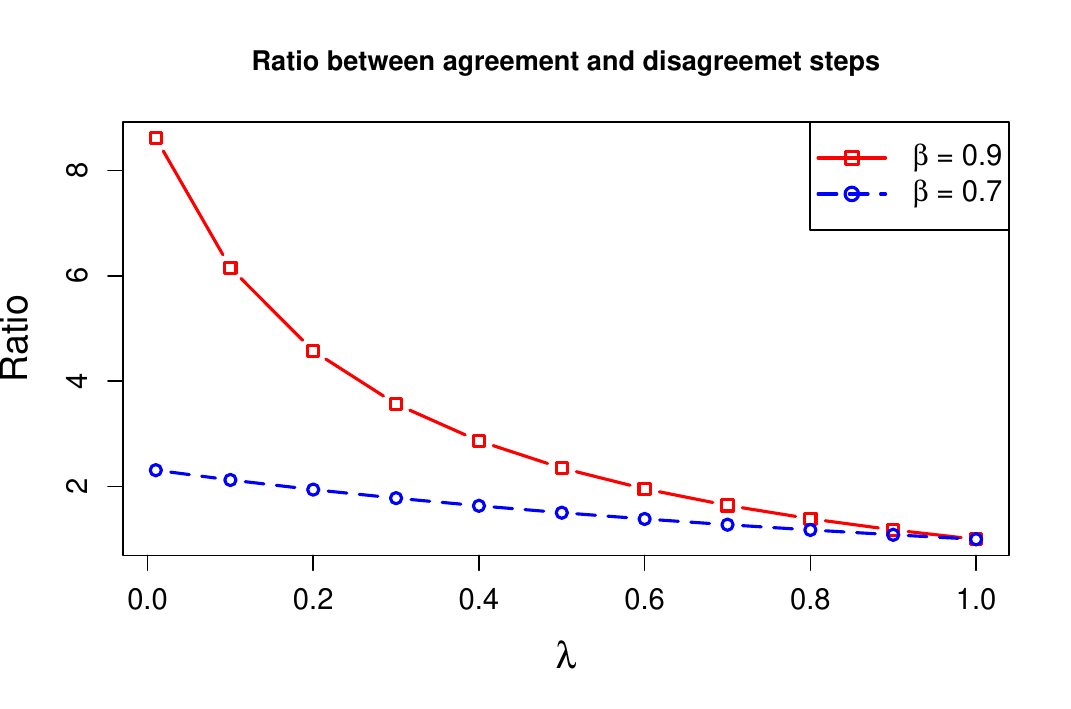}
	\end{tabular}
	\caption{ Step sizes as a function of the estimate proportion $\lambda$ of honest agents among those each agent is biased against. {\it Top panel:} Size of the steps for agreement, $SA$, versus disagreement, $SD$ when it is believed dishonest agents lie with probability $\beta=0.9$. {\it Bottom Panel:} Ratio $s=SA/SD$ for two values of $\beta$.
	}\label{fig:stepsizes}
\end{figure}

Figure~\ref{fig:stepsizes} shows how the step sizes change as a function of the estimated proportion $\lambda$ of honest agents among those each agent is biased against. In the top panel, we can see both the step sizes when the neighbor agrees with the opinion favored by the bias, $SA$, and when the neighbor disagrees with it, $SD$.  Notice that when $\lambda$ tends to 1.0, both steps become equal. That corresponds to the scenario where everyone is honest; we recover the original CODA model with identical steps. The apparent equality when $\lambda$ tends to zero is not real and is only a product of visualizing minimal steps. Indeed, the bottom panel shows that the ratio between the steps,  $s=SA/SD$, increases continuously as $\lambda$ gets close to zero.

The changes in the step size are obviously not of the same size. If we want to normalize the steps, as in CODA, we can choose either $SA$ or $SD$ as the step we make equal to 1.0. We will initially make, for the implementations, the smaller $SD=1.0$. And, as a simplification, instead of carrying dependencies on $\lambda$, $\alpha$, and $\beta$, we will simply assume there is a ratio $s$ such that $SA = s SD$. That is, we will assume that disagreement with the bias corresponds to a step size of 1.0 and agreement with the bias to a step size of $s\geq 1.0$, where $s=1.0$ corresponds to the situation where there is no bias. Finally, we have elementary update rules we can implement given by

\begin{equation}\label{eq:directupdate}
	\nu_i(t+1)=  \nu_i(t) + 
	\begin{cases}
		sign(\nu_j) & \text{if neighbor $j$ disagrees with $i$ bias,}\\
		s*sign(\nu_j) & \text{if neighbor $j$ agrees with $i$ bias.}
	\end{cases}
\end{equation}

 \subsubsection{Conservatism bias}

As a related example, let us introduce the effect called conservatism  \cite{edwards68}, where people change their opinions less than they should. That can be quickly introduced using a mental model where the agent thinks there is a chance the data is reliable and a chance that the information is only non-informative noise. When that happens, it is only natural that the update's size will be much smaller. The larger the chance associated with noisy data, the smaller the update's size.

We have the same mathematical problem we had with the direct bias, except that now, instead of believing that defenders of a specific side might be lying, the agents think that defenders of the side who disagree with them might be dishonest. Suppose an agent changes its opinion from $A$ to $B$. In that case, it will change its assessment of where there might be dishonesty from $B$ to $A$. That means we have very similar rules as those in Equation \ref{eq:directupdate}. However, the bias is always the same as $i$ opinion, so the rules depend directly on $sign(\nu_i)$. That is, we have
  
 \begin{equation}\label{eq:directupdateconserva}
 	\nu_i(t+1)=  \nu_i(t) + 
 	\begin{cases}
 		sign(\nu_j) & \text{if $sign(\nu_j)\neq sign(\nu_i) $,}\\
 		s*sign(\nu_j) & \text{if $sign(\nu_j) = sign(\nu_i) $.}
 	\end{cases}
 \end{equation}

 \section{Results}
 
 To observe how the system might evolve under each rule, we implemented the models using the R software environment \cite{Rsoftware}. The code is available at \url{https://www.comses.net/codebase-release/d4ab2a25-4233-4e6e-a8c5-a3b919cfd6e2/}

All cases shown here correspond to an initial neighborhood of agents defined as a square, bi-dimensional lattice with $40^2$ agents, no periodic conditions, and second-level neighbors. As commented below, in some situations, we let the network evolve into a polarized situation before allowing opinions to change using this polarized and quenched network. Once that initial network, lattice, or rearranged, is established, agents interact, observing the choice of one neighbor and updating their own opinion based on that observation, according to the update rules of each case. There was an average number of 50 interactions per agent.  For all simulations, initial opinions were drawn from a uniform continuous distribution on the range $-2.0\leq \nu_i \leq +2.0$. All the results for the distribution of opinions correspond to averages over 20 realizations of each case.

	\subsubsection{Simulating a direct bias} 

\begin{figure}
	\centering
	\begin{tabular}{c}
		\includegraphics[width=0.85\textwidth]{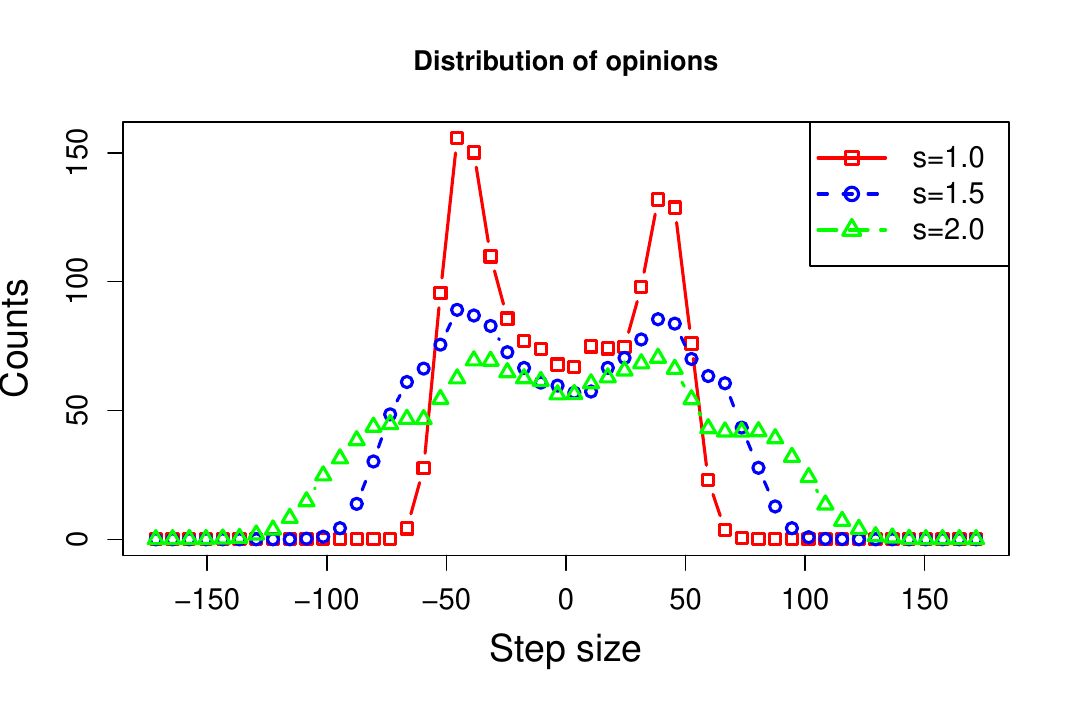}\\
		\includegraphics[width=0.85\textwidth]{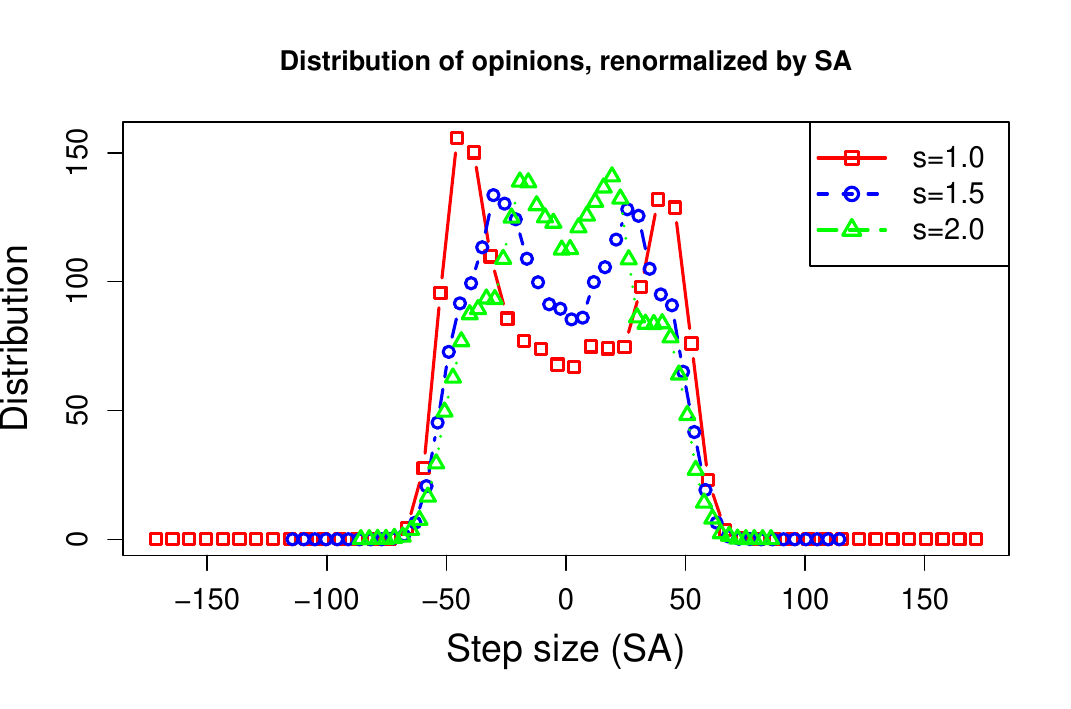}
	\end{tabular}
	\caption{ Distribution of opinions after  {\it Top panel:} Opinions measured as disagreement steps. {\it Bottom Panel:} Opinions renormalized to agreement steps.
	}\label{fig:nsize40nei2nrun50nav20norewire}
\end{figure}

Results for the distribution of opinions in the direct bias case, with no initial rewiring, for three distinct values of the ratio $s$ can be seen in Figure~\ref{fig:nsize40nei2nrun50nav20norewire}. Each curve corresponds to a different ratio $s$ between the agreement and the disagreement step. The top graphic shows how extreme the opinion is measured in the number of disagreement steps (following the algorithm implementation). The bottom picture shows the renormalized distribution if we measure opinions in terms of agreement steps.

It is easy to observe in the upper figure, with the distribution measured in disagreement steps, that as $s$ increases, the opinions spread further away from the central position. That suggests that the opinions become more extreme. While the peaks associated with the more extreme opinions become softer (and seem to disappear for $s=2.0$), that happens because the existing extremists get distributed over an extensive range of even stronger opinions.

A different picture emerges if we look at a renormalized step size, using the agreement step as unit, $SA=1$, instead of the implemented disagreement step. The distribution of opinions for such a case can be seen in the bottom graphic in Figure~\ref{fig:nsize40nei2nrun50nav20norewire}. What we observe, in this case, is that when we measure the strength of opinions using $SA$ as the measuring unit, the distributions become much more similar. As $s$ increases, we observe an increase in the number of agents around the weaker opinions and smaller peaks of extremists.

The apparently contradictory conclusions we can arrive at by looking at only one graphic are another example of the problem with adequately defining what an extremist is \cite{Martins2022a}. In this model, we actually have two different ways to define extremism. One of those definitions would arise if we were to transform back the number of steps into probability values by inverting the renormalizations and transformations of variables. It is worth noticing that, as can be seen in the upper graphic in Figure~\ref{fig:stepsizes}, both step sizes, for agreement and disagreement, become smaller when bias is introduced when compared to the unbiased case $s=1.0$. That would lead to the strange conclusion that when agents think others might be biased, their final opinion becomes less extreme. That makes sense if we only care about the agents' confidence in absolute terms. After all, other agents become less reliable. Their information should convince less.

But there is another definition of extremism that is also natural and reasonable. That definition comes from asking how easy (or hard) it would be to change the choice of a specific agent. This does correspond to the number of steps away from the central opinion. More precisely, since it is necessary to move to the opposite view to change one's choice, using the disagreement step as the unit is the best choice. 

That would be the case if we were studying results from the conservatism bias, as we will do in Subsection~\ref{sec:conservatismresults}. Here, however, bias was fixed, corresponding to the initial choice of the agent. And that means that there is an essential proportion of agents whose biases do not conform to their opinions. While the proportion of agreement between bias and final opinion increases with $s$ (observed averages were 57.7\% for $s=1$, 66.4\% for $s=1.5$, and 74.5\% for $s=2.0$), we never obtained a perfect match between bias and opinion. Even at $s=2.0$, about 1 in 4 agents would move towards the opposite choice using agreement steps.

That might seem unrealistic. While there may be a bias toward initial opinions, people usually defend their current position. What size of step humans would use when returning to a previously held belief is an interesting question, but that is beyond the scope of this paper. On the other hand, confirmation bias is 
described not as an agreement with initial views but as the tendency to look for sources of information that agree with our current beliefs.

That bias was already studied before, with a network that evolved simultaneously with the opinion updates \cite{Martins2019}. As agents stopped interacting with those they disagreed with, we had a case of traditional confirmation bias. That also means that the results we have analyzed so far correspond to a direct bias independent of opinions. Real or not, it was introduced here as an exercise and as an example that we can model different modes of thinking using Bayesian tools, regardless of if they correspond to reality.

\subsection{Rewiring the network}

Of course, we want to explore more realistic cases. We will do that in two steps, first, by getting closer to a confirmation bias by introducing rewired networks, and second, by implementing the model of conservatism bias, as defined in Equation~\ref{eq:directupdateconserva}.

Here, we will follow the rewiring algorithm previously used to study the simultaneous evolution of networks and opinions to generate an initial, quenched network before opinion updates start. At each step, the algorithm tries to destroy an existing link between two agents (1 and 2) and create a new one between two other agents (3 and 4). The decision of whether to accept that change depends on the euclidean distance between agents 1 and 2, $d_{12}$, and agents 3 and 4, $d_{34}$, measured by considering a coordinate system over the square lattice where first neighbor distances correspond to 1. That way, there is a tendency to preserve the initial square lattice. And we also use a term that makes it more likely to accept the change when the old link was between disagreeing agents and the new one is between agreeing agents. That is, each rewiring is accepted with probability

\begin{equation}\label{eq:probtransdist}
	P=\exp(-\Delta H )= \exp[-\beta(d_{34}-d_{12} - J (\sigma_3 \sigma_4 - \sigma_1 \sigma_2 ))],
\end{equation}

where $J$ is the relative importance between physical proximity and opinion agreement. Notice that if we want a reasonable chance that distant agents will connect, $J$ should be comparable with the initial side of the square network. 

\begin{figure}
	\centering
	\begin{tabular}{c}
		\includegraphics[width=0.95\textwidth]{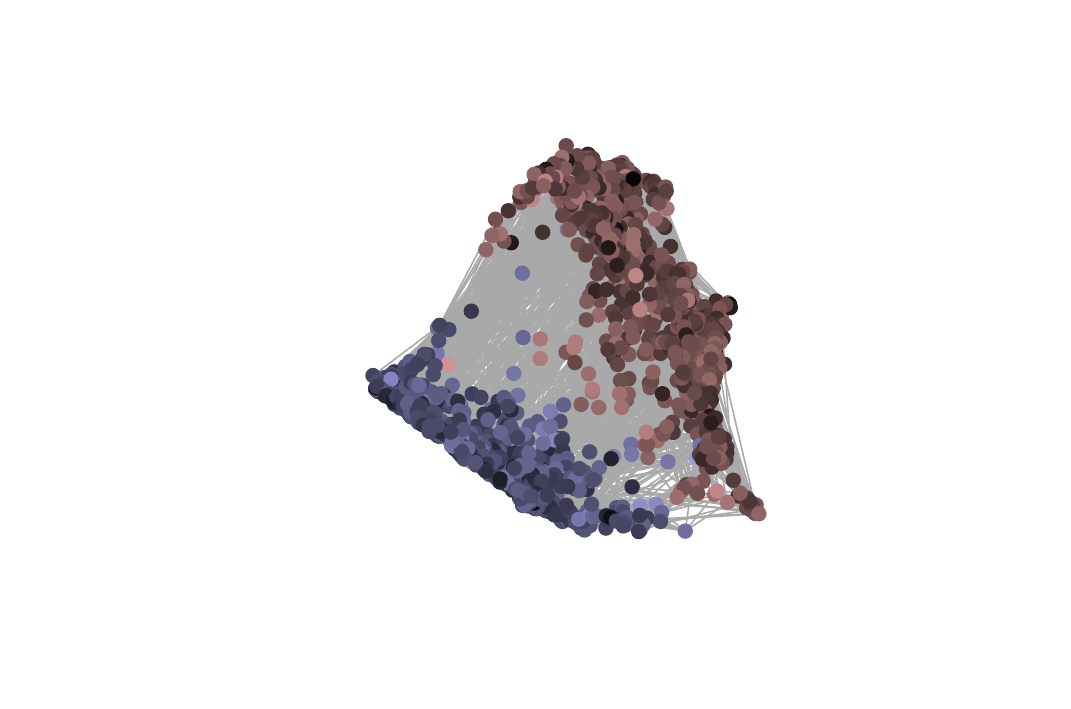}
	\end{tabular}
	\caption{Typical network formed after applying the rewiring algorithm with an average of 20 rewirings per agent, $\beta=1.0$, and $J=20$. Red and blue colors correspond to the two choices, and darker hues show a stronger final opinion obtained after the opinion update phase.}\label{fig:typical}
\end{figure}

Figure~\ref{fig:typical} shows a typical case, as an example, of how an initially square lattice is altered after an average of 20 rewirings per agent, $\beta=1.0$, and $J=20$. In the figure, red and blue colors correspond to the two choices, and darker hues show a stronger final opinion, obtained after the opinion update phase. As the network did not change during the opinion update phase, its final shape after implementing the rule defined in Equation~\ref{eq:probtransdist} is preserved.

\begin{figure}
	\centering
	\begin{tabular}{c}
		\includegraphics[width=0.85\textwidth]{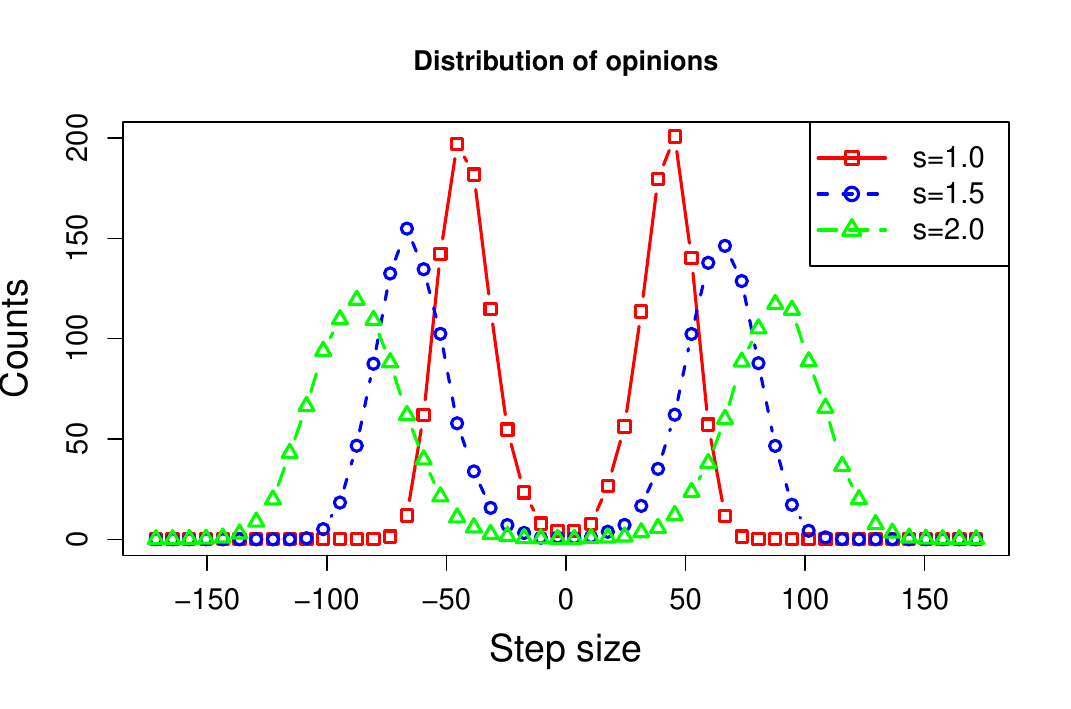}\\
		\includegraphics[width=0.85\textwidth]{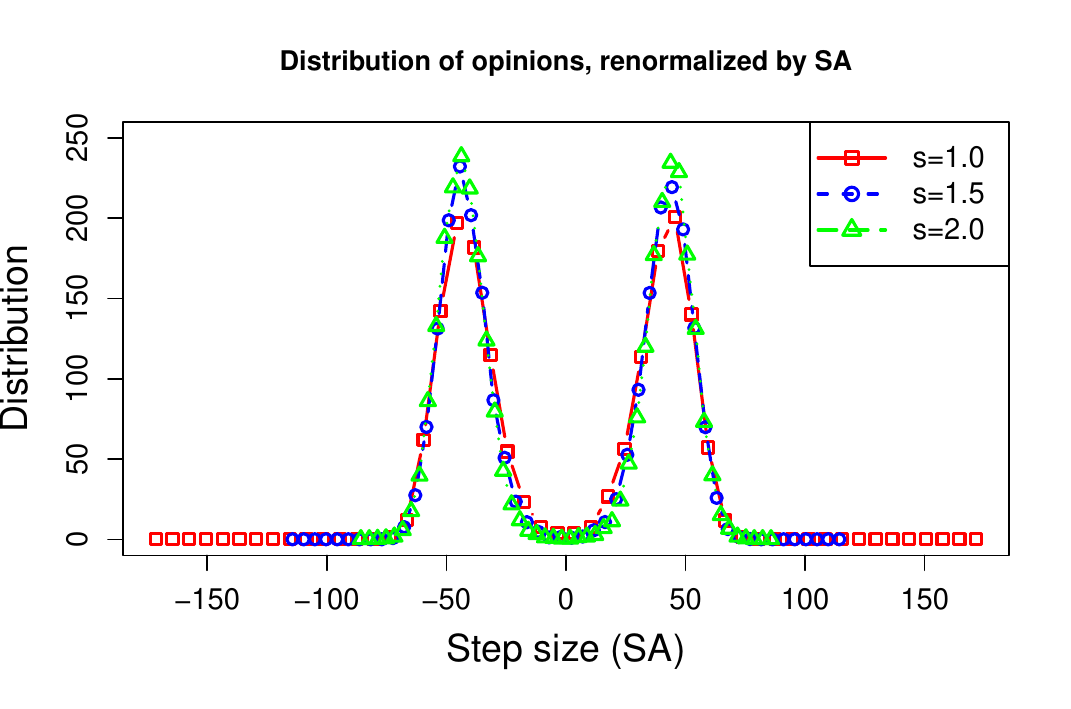}
	\end{tabular}
	\caption{ Distribution of opinions after  {\it Top panel:} Opinions measured as disagreement steps. {\it Bottom Panel:} Opinions renormalized to agreement steps.
	}\label{fig:nsize40nei2nrun50nav20ini20}
\end{figure}

Figure~\ref{fig:nsize40nei2nrun50nav20ini20} shows the distribution of opinions when we use the direct bias rules after obtaining a quenched network using the parameters to generate networks similar to those in Figure~\ref{fig:typical}. We can see significant differences between the results with no initial rewiring (Figure~\ref{fig:nsize40nei2nrun50nav20norewire}) and the new ones (Figure~\ref{fig:nsize40nei2nrun50nav20ini20}). With the initial rewiring, we find almost no moderates in the final opinions, and the extremist peaks, when we use the disagreement step, $SD$, for normalization, move to even stronger values as $s$ increases. That dislocation also corresponds to smaller peaks distributed over a more extensive range. That, however, is an artifact of using $SD$. When renormalized to $SA$ size steps, we see that the three curves for different values of $s$ match almost perfectly. That happens because, with the implemented rewiring, most interactions happen with agents who already share the same opinions.

However, the problem of defining extremism under these circumstances becomes much more straightforward. We no longer have a situation where the agent bias does not agree with its opinion. Indeed, the observed averages for the proportion of agreement between bias and opinions were 99.8\% for $s=1$, 100\% for $s=1.5$, and 100\% for $s=2.0$. That means that to change position, all agents would move one disagreement step at a time. While observing how the curves match when we measure opinions in agreement steps is interesting, the disagreement step case is more informative. And here, as expected, the more bias one introduces, the harder it becomes for all agents to change their opinions.

 	\subsubsection{Simulating the conservatism bias}\label{sec:conservatismresults}
 	
 	We also implemented the conservatism model defined in Equation~\ref{eq:directupdateconserva} using the same parameter values we used in the previous cases. As in this scenario, there is no distinction between opinion and bias, the simulations presented here do not include an initial rewiring phase to guarantee that most agents are aligned with their own biases.

\begin{figure}
	\centering
	\begin{tabular}{c}
		\includegraphics[width=0.85\textwidth]{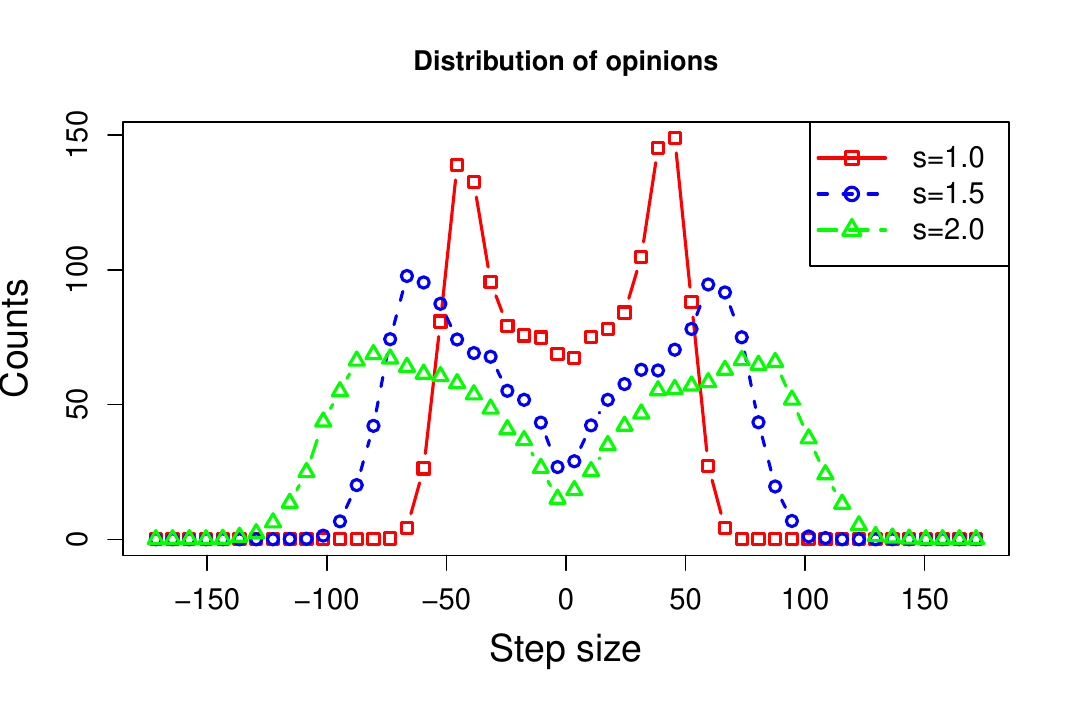}\\
		\includegraphics[width=0.85\textwidth]{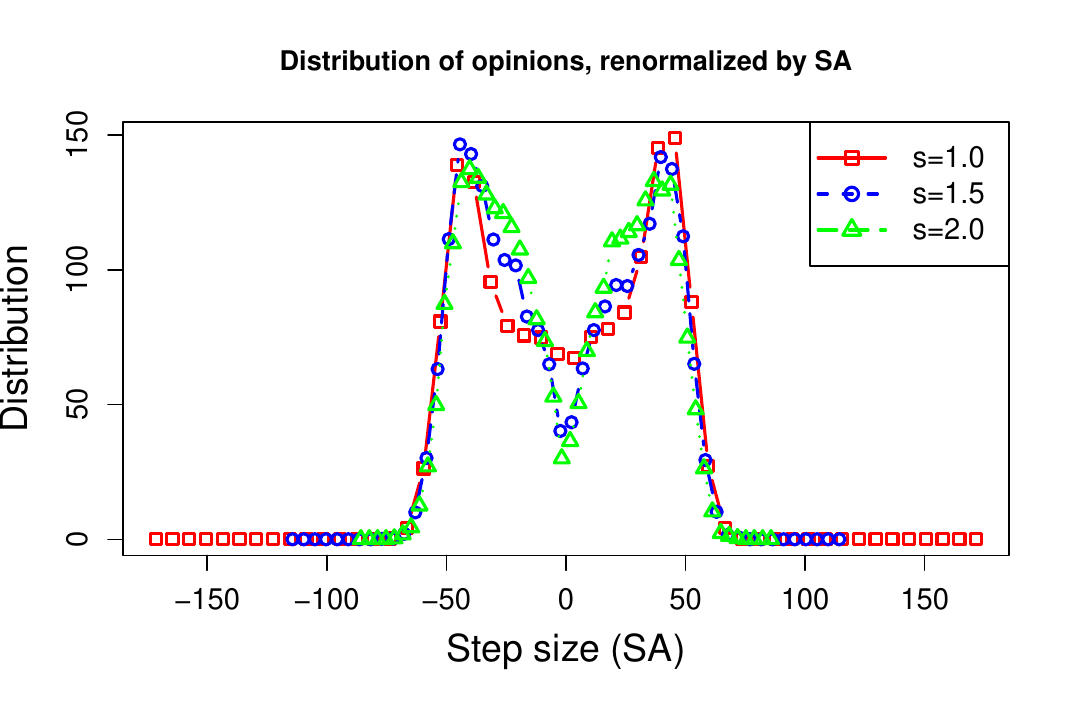}
	\end{tabular}
	\caption{ Distribution of opinions after  {\it Top panel:} Opinions measured as disagreement steps. {\it Bottom Panel:} Opinions renormalized to agreement steps.
	}\label{fig:nsize40nei2nrun50nav20norewire_conservatism}
\end{figure}

Figure~\ref{fig:nsize40nei2nrun50nav20norewire_conservatism} shows the distribution of opinions as disagreement steps (top panel) and agreement steps (bottom panel). As we have discussed, in this case, the measure that tells us how hard it would be for an agent to change its choice is associated with the disagreement steps. And we can see at the top figure that, as we had observed with the direct bias case with no initial rewiring, the peaks of extreme opinions move to more distant values and become less pronounced. That happens because, contrary to the rewired situation where there were few links between disagreeing agents, we still have more extensive borders where we can find agents whose neighbors have distinct choices. Despite that, moderate agents, close to zero, become rarer as the conservatism bias increases. While not as relevant for understanding how extreme opinions are in this model, the graphic in the bottom figure, renormalized for agreement steps of size one, is still interesting. It shows the same tendency of preserving the general shape for different values of $s$ we observed before, except that it is clear that when no conservatism is expected ($s=1.0$), it still shows more agents in the moderate region.

 \section{Conclusion}
 
Approximating the complete but impossible Bayesian rules provides a reasonable description of human behavior, especially when we account for individuals' imperfect trust in others and their tendency to reason in a motivated manner. By modeling these approximations, we can create realistic models of human behavior, as demonstrated in this paper. Our analysis focuses on introducing variations to the CODA model where agents exhibit biases in favor of a particular opinion. Through our investigation, we find that conservatism implementation, where the agents distrust information that goes against their beliefs, results in more extreme opinions and a greater resistance to change.

Furthermore, our research examines how to obtain update rules from assumptions about the agents' mental models, using both previously published cases and our new examples. Another crucial aspect of using Bayesian-inspired rules is that it allows for a clearer understanding of the relationship between distinct models. By exploring which assumptions lead to our current opinion models, we gain insight into their interrelatedness and identify situations where each model might be more applicable. Overall, our work sheds light on the potential of Bayesian-inspired modeling to offer a more nuanced description of agent behavior and its impact on opinion dynamics.
 
 	\section{Acknowledgments}
 
 This work was supported by the Funda\c{c}\~ao de Amparo a Pesquisa do Estado de S\~ao Paulo (FAPESP) under grant  2019/26987-2.

%
\bibliographystyle{unsrt}
\bibliography{biblio}

\end{document}